\documentclass[aps,prb,amsmath,amssymb,twocolumn]{revtex4}
\usepackage{graphicx}
\usepackage{hyperref}

\begin{document}

\title{Equilibrium edge spin currents in two-dimensional electron systems with spin-orbit interaction}

\author{Vladimir~A.~Sablikov, Aleksei~A.~Sukhanov, and Yurii~Ya.~Tkach}

\affiliation{Kotel'nikov Institute of Radio Engineering and Electronics,
Russian Academy of Sciences, Fryazino, Moscow District, 141190,
Russia}

\begin{abstract}
We find a specific mechanism of background spin currents in two-dimensional electron systems with spatially nonuniform spin-orbit interaction (SOI) at thermodynamic equilibrium, in particular, in the systems consisting of regions with and without SOI. The spin density flows tangentially to the boundary in the normal and SOI regions within a layer of the width determined by the SOI strength. The density of these edge currents significantly exceeds the bulk spin current density. The spin current is polarized normally to the boundary. The spatial distribution of equilibrium spin currents is investigated in detail in the case of a step-like variation of the SOI strength and potential at the interface.

\end{abstract}

\maketitle

The generation and manipulation of electron spin polarization in semiconductor nanostructures solely by means of electric fields has attracted significant attention as an interesting manifestation of spin-orbit interaction (SOI) as well as a valuable capability for spintronic devices.~\cite{Wolf,Zutic} A key point in solving this problem is searching for mechanisms of effective generation of purely spin currents. In this connection a great deal of interest was paid recently to the dissipationless spin currents in semiconductors with SOI~\cite{Murakami,Sinova} and particularly, to the spin currents in two-dimensional (2D) electron systems under the thermodynamic equilibrium.~\cite{Rashba1,Pareek} The existence of equilibrium background spin currents was first pointed out by Rashba~\cite{Rashba1} for an infinite 2D electron gas (2DEG) with SOI. This phenomenon has arisen a wide discussion of the definition of spin currents~\cite{Rashba2,Burkov,Shi,Sonin1,Sun} and the possibility to observe the equilibrium spin currents~\cite{Sun,Sonin2}. The definition of the spin currents is known to be somewhat arbitrary since the spin is not conserved in the presence of SOI. The discussion has led to the evident conclusion that observable effects do not depend on any definitions. One can use the standard definition of spin current to calculate correctly measurable values. Sonin~\cite{Sonin2} proposed to detect the equilibrium spin current by measuring the mechanical torques near the edges of the SOI medium. Sun et al~\cite{Sun} argued that the equilibrium spin current can be interpreted as a persistent current, and considered the possibility to determine this current in a ring device by measuring an electric field it generates.

In this paper we draw attention to another aspect of the spin current problem. We study equilibrium spin currents in a 2D electron system with nonuniform SOI. We were initially motivated by a question of whether the equilibrium spin currents penetrate from a 2DEG with SOI into normal 2DEG. This problem is interesting by two reasons: (i) if the spin current could penetrate into a normal 2DEG, it could be a measure of the spin current in SOI medium since in a normal 2DEG the spin current is unambiguously defined;~\cite{n1} (ii) many realistic 2D structures usually contain regions in which Rashba SOI is induced by a normal electric field produced by gates forming mesoscopic structures or by built-in charges. In particular, the SOI apparently exists at edges of constrictions in 2DEG and may affect the spin-dependent electron transport observed in experiments~\cite{Rokhinson}.

We study the equilibrium spin currents in a 2D electron system consisting of regions with and without SOI to come to unexpected conclusion that very effective mechanism of the spin current generation acts near the boundary. This mechanism generates the equilibrium edge spin current of significantly higher density than the bulk spin current in an infinite 2DEG with SOI. The edge spin current density is proportional to the SOI constant $\alpha$, while the bulk current is proportional to $\alpha ^3$. To be more specific, the edge and bulk currents are, respectively, proportional to $ak_F^2$ and $a^3$, where $a$ is the SOI wave vector and $k_F$ is the Fermi wave vector. The difference in the magnitudes of these currents is caused mainly by the fact that the edge spin current is produced by electrons in the whole energy range of occupied states, while the bulk spin current is generated in an energy layer, the width of which equals to the characteristic SOI energy. A remarkable result is that the equilibrium edge spin currents exist in the normal 2DEG where the spin current is strictly defined. 

\begin{figure}[h,t]
\includegraphics[width=0.9\linewidth]{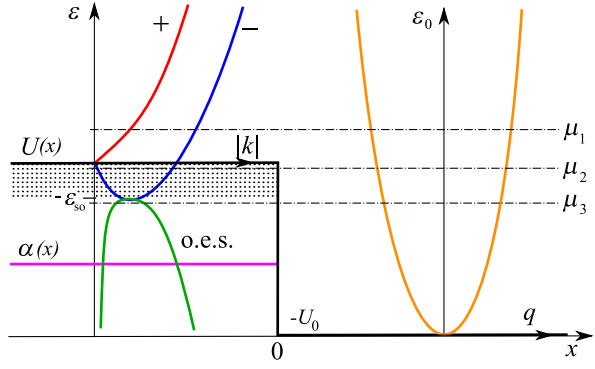}
\caption{(Color online). The energy diagram of the normal-SOI structure.  $\varepsilon (|k|)$ denote electron spectra in the SOI region; (+) and (-) are upper and lower states with propagating and decaying evanescent modes; (o.e.s.) is ``oscillating'' evanescent modes. Dotted is the energy layer, in which the equilibrium spin currents are generated in the bulk of the SOI reservoir. $\varepsilon_0(q)$ is the spectrum in the normal region.}
\label{sharp_junction}
\end{figure}

Let us consider the equilibrium spin currents in a 2D system consisting of semi-infinite regions with and without Rashba SOI (Fig.~\ref{sharp_junction}). The SOI is described by a step function $\alpha(x)=\alpha\Theta(-x)$. The potential energy $U(x)$ is also step function, the potential in the SOI region being higher than in the normal one. This arrangement of the potential shape allows one to study effects of both propagating and evanescent modes in the SOI region on the spin state in the normal region.

First, we briefly describe the spectrum of electron states in the SOI region. There are three types of modes studied in detail in Ref.~\onlinecite{Sablikov}. The eigenfunctions are
\begin{equation}
\psi_{\mathbf{k},\lambda}=C \binom{\chi_{\lambda}(\mathbf{k})}{1}e^{i(k_x x+k_y y)}\,,
\label{psi1}
\end{equation}
where $\lambda$ stands for spin states, $\mathbf{k}=(k_x,k_y)$, and $C$ is a constant. Since in the case being considered, $k_y$ is a conserving value, we will drop this index in the notations of wave functions. The characteristic wave vector and energy of the SOI are $a=m\alpha/\hbar^2$ and $\varepsilon_{so}=\hbar^2a^2/(2 m)$.

In the energy range $\varepsilon >-\varepsilon_{so}$, there are two modes: 

(i) propagating modes $|\lambda,k'_x\rangle $, with $k_x$ being real, $k_x=k'_x$; $\lambda=\pm$; and $\chi_{\lambda}(\mathbf{k})=\lambda (k_y+i k'_x)/k$;

(ii) ``decaying'' evanescent modes $|\lambda,k''_x\rangle $, with $k_x$ being imaginary, $k_x=i k''_x$; $\lambda=\pm$; and $\chi_{\lambda}(\mathbf{k})=\lambda (k_y- k''_x)/k$.

The energy of both modes above is
\begin{equation}
 \varepsilon_{\lambda,\mathbf{k}}=\dfrac{\hbar^2}{2m}\left[-a^2+\left(a+\lambda k\right)^2\right]\,,
\label{dispersion_1_2}
\end{equation}
where $k=(k_x^2+k_y^2)^{1/2}$. At a given energy $\varepsilon >0$, the wave vector $k$ takes two values $k_1$ and $k_2$ corresponding to the states with positive and negative chiralities ($\lambda=\pm$):
\begin{equation}
\label{k12}
k_{1,2}=-\lambda a+\sqrt{2m\varepsilon/\hbar^2+a^2}\,.
\end{equation}

In what follows one needs to know which states exist at a given energy. For $\varepsilon >0$, three cases are possible. If $|k_y|<k_1$, there are two pairs of propagating states $|\pm k'_{1x}\rangle$ and $|\pm k'_{2x}\rangle$. If $k_1<|k_y|<k_2$, the states with $\lambda=+1$ are ``decaying'' evanescent $|\pm k''_{1x}\rangle$ while the states with $\lambda=-1$ remain propagating $|\pm k'_{2x}\rangle$. If $k_2<|k_y|$, both states with $\lambda=\pm 1$ are ``decaying'' evanescent $|\pm k''_{1x}\rangle$ and $|\pm k''_{2x}\rangle$.

In the range $-\varepsilon_{so}<\varepsilon<0$, the states with positive chirality disappear and all four states have $\lambda=-1$. In this case there are two characteristic wave vectors
\begin{equation}
\label{k_neg_ch}
\tilde{k}_{1,2}=a \mp \sqrt{2m\varepsilon/\hbar^2+a^2}\,.
\end{equation}
If $|k_y|<\tilde{k}_1$, two pairs of propagating states $|\pm \tilde{k}'_{1x}\rangle$ and $|\pm \tilde{k}'_{2x}\rangle$ exist. If $\tilde{k}_1<|k_y|<\tilde{k}_2$, a pair of the states are ``decaying'' evanescent $|\pm \tilde{k}''_{1x}\rangle$ and another pair is propagating $|\pm \tilde{k}'_{2x}\rangle$. If $\tilde{k}_2<|k_y|$ all states are ``decaying'' evanescent, $|\pm \tilde{k}''_{1x}\rangle$ and $|\pm \tilde{k}''_{2x}\rangle$. It is worth noting that it is just the energy layer that generates the equilibrium spin currents in an infinite 2DEG with SOI.\cite{Rashba1}

In the energy range $\varepsilon <-\varepsilon_{so}$, the electron states are ``oscillating'' evanescent $|k'_x,k''_x\rangle $. They are described by a complex wave vector $k_x=k'_x+i k''_x$,
\begin{equation}
\label{dispersion_3}
\begin{split} 
\varepsilon_{\lambda,\mathbf{k}}=-\dfrac{\hbar^2}{2m}\left(a^2+\dfrac{{k'_x}^2{k''_x}^2}{a^2}\right)\,,\\
\chi(\mathbf{k})=-\dfrac{a(k_y-k''_x+ik'_x)}{a^2+ik'_xk''_x}\,.
\end{split}
\end{equation}
Here $k'_x,k''_x$, and $k_y$ are coupled by an equation described in Ref.~\onlinecite{Sablikov}. At a given energy there are four complex vectors $k_x$ corresponding to different signs of $k'_x$ and $k''_x$.

Note that in the bulk of the SOI reservoir solely the propagating states $|\lambda,k'_x,k_y\rangle $ exist. Evanescent states can be realized only near the interface.

In the normal reservoir the electron states $|\lambda,q,k_y\rangle$ are propagating. They are degenerate by the spin:
\begin{equation}
 |\lambda,q\rangle=\dfrac{1}{\sqrt{2}}\binom{\lambda (k_y+iq)/(k_y^2+q^2)^{1/2}}{1}e^{i(q x+k_y y)}\,,
\label{lambda_q}
\end{equation}
where $q$ is the $x$ component of the wave vector.

Now let us turn to the normal-SOI structure. The wave functions can be easily written for a region of a width shorter than electron mean free path near the interface. The electron states are formed by the waves coming from the left and right reservoirs. They partially transmit the interface and reflect from it. Correspondingly there are four types of wave functions of the form:\\
(i) for right moving electrons
\begin{equation}
 \psi^{(R)}_{s,\mathbf{k}}\!=\!\! \left\{
\begin{aligned}
&|s,k_x\rangle + \sum_{s'}r^R_{s,s'}|s',-k_x\rangle \,, & \quad x<0\,,\\
&\sum_{s'}t^R_{s,s'}|s',q\rangle \,, & \quad x>0\,;
\end{aligned}
\right.
\label{Rwave}
\end{equation}
(ii) for left moving electrons
\begin{equation}
 \psi^{(L)}_{s,\mathbf{k}}\!=\!\! \left\{
\begin{aligned}
&|s,q\rangle + \sum_{s'}r^L_{s,s'}|s',-q\rangle \,, &\quad x>0\,,\\
&\sum_{s'}t^L_{s,s'}|s',q\rangle \,, &\quad x<0\,.
\end{aligned}
\right.
\label{Lwave}
\end{equation}
Here $r^{R/L}_{s,s'}$ and $t^{R/L}_{s,s'}$ are reflection and transmission matrices. Index $s=1,2$ numerates the states in the SOI and normal regions, which participate in the scattering process at the interface. For the state vectors in the normal region, the index $s=1,2$ corresponds to the chirality $\lambda=\pm$ in Eq.~(\ref{lambda_q}). For states in the SOI region, the situation is more complicated. In the energy region $\varepsilon>0$, the index $s=1,2$ also matches with the chirality $\lambda=\pm$. In the energy layer $-\varepsilon_{so}<\varepsilon <0$, where $\lambda=-1$, the index $s=1,2$ labels two roots defined by Eq.~(\ref{k_neg_ch}). For $\varepsilon <-\varepsilon_{so}$ the index $s$ labels two states with negative $k''_x$ and different signs of $k'_x$: $s=1$ corresponds to $k'_x>0$, and $s=2$ does to $k'_x<0$.

The reflection and transmission matrices are determined by matching the wave functions (\ref{Rwave}) and (\ref{Lwave}) at the interface using the standard boundary conditions~\cite{Sablikov}:
\begin{equation}
 \psi|^{+0}_{-0}=0, \qquad \left[\dfrac{d\psi}{dx}-(i a \hat{\sigma}_y-\beta k_y \hat{\sigma}_z)\psi \right]^{+0}_{-0}=0\,,
\end{equation}
where $\beta$ takes into account the Rashba SOI caused by an in-plane gradient of the potential at the interface.

The spin current is defined as
\begin{equation}
 \mathcal{J}_i^{\gamma}(r,s,\mathbf{k})=\dfrac{\hbar}{4}\langle \psi_{s,\mathbf{k}}^{(r)}|\hat{\sigma}_{\gamma}\hat{v}_i+\hat{v}_i\hat{\sigma}_{\gamma}|\psi_{s,\mathbf{k}}^{(r)} \rangle\,,
\label{Js}
\end{equation}
where $r$ labels the right and left moving states and $\gamma$ stands for the polarization. This definition is strictly validated for the normal region of the structure under consideration, where Eq.~(\ref{Js}) follows from the spin conservation law. In the SOI region the spin current is not conserved and therefore can not be unambiguously defined. Equation~(\ref{Js}) represents the commonly accepted and physically appealing definition of spin current.~\cite{Murakami,Sinova,Rashba1,Pareek,Rashba2,Burkov,Sonin1,Sun} We emphasize that the spin currents in the normal region do not depend on the definition of the spin current in the SOI region.

The total current is a sum over all electron states below the chemical potential $\mu$, 
\begin{equation}
 \mathcal{J}_i^{\gamma}(\mu)\!=\sum_r^{R,L}\sum_s^{1,2}\!\!\int \limits_{-U_0}^{\mu}\!\! d\varepsilon \!\! \int \!\!dk_y D^{(r)}_s\!(\varepsilon,k_y) \mathcal{J}_i^{\gamma}(r,s,\mathbf{k}),
\end{equation} 
where $D^{(r)}_s\!(\varepsilon,k_y)$ is the density of states for waves incident on the interface from the left or right reservoirs.

\begin{figure}[h,t]
\includegraphics[width=0.9\linewidth]{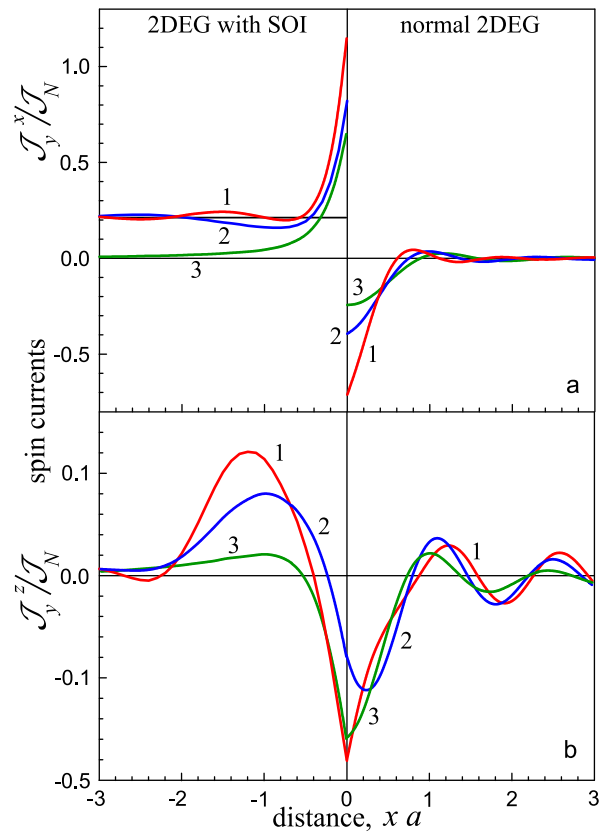}
\caption{(Color online). The distribution of spin current directed tangentially with spin polarization in (a) $x$ direction and (b) $z$ direction. Curves 1, 2, and 3 correspond to the chemical potentials $\mu_1$, $\mu_2$, and $\mu_3$ shown in Fig.~\ref{sharp_junction}. Thin horizontal line shows the equilibrium spin current density in infinite 2DEG with SOI. The current is normalized to $\mathcal{J}_N=\hbar^2a^3/4m$. Calculations are carried out for $U=6\varepsilon_{so}$ and $\beta=-0.01$.}
\label{Jyxz}
\end{figure}

Calculations show that in the normal region all components of the spin currents $\mathcal{J}_x^{\gamma}$ directed normally to the interface are absent, though in an infinite 2DEG with SOI the spin current $\mathcal{J}_x^{y}$ is nonzero.~\cite{Rashba1} Hence, the normal component does not come outside the SOI medium, in any case through a smooth border line we are considering here. The tangential component with tangential polarization $\mathcal{J}_y^{y}$ is also absent. Only nonzero components are $\mathcal{J}_y^{x}$ and $\mathcal{J}_y^{z}$. Thus in the normal region spins flow tangentially to the interface and are polarized normally to the interface. The spatial distribution of the spin currents is presented in Fig.~\ref{Jyxz}. In the normal region the spin current is located at a distance of the order of $1/a$ and after some oscillations vanishes in the bulk.

In the SOI region the spatial distribution of the tangential spin currents essentially depends on the chemical potential. If $\mu<-\varepsilon_{so}$ the spin current is caused by the ``oscillating'' evanescent states and therefore vanishes with distance. In the case $\mu>-\varepsilon_{so}$, the $z$-polarized current vanishes with the distance while the $x$-polarized current goes to a finite value coinciding with the spin current in infinite system.~\cite{Rashba1} The normal components $\mathcal{J}_x^{\gamma}$ turn to zero at the interface for all polarizations. Notice that the spin density is zero everywhere.

Of most importance are the following facts: (i) near the interface the spin current is much higher than the maximum of the current attainable in an infinite system; (ii) the near-boundary spin current is generated by electrons with any energy in contrast to the case of infinite system where the equilibrium spin current is generated in the energy layer $-\varepsilon_{so}<\varepsilon<0$. These facts evidence that near the interface another mechanism of the equilibrium spin current acts, which is caused by the spin-dependent scattering of electrons by the interface.~\cite{n2}

We demonstrate this mechanism by considering a simplified situation where an uniform 2DEG contains a region of the SOI in the form of a strip lying along the $y$ axis. The SOI strength $\alpha$ is a continuous function of $x$ localized at a distance $2l$. The Schr\"odinger equation is
\begin{equation}
 -\dfrac{\hbar^2}{2m}\dfrac{d^2\psi}{dx^2}+\hat{V}(x)\psi=\left(\varepsilon-\dfrac{\hbar^2k_y^2}{2m}\right)\psi\,,
\end{equation}
where 
\begin{equation}
\hat{V}(x)=\alpha(x)\left(k_y\hat{\sigma}_x+i\hat{\sigma}_y\dfrac{d}{dx}\right)+ \dfrac{i}{2}\dfrac{d\alpha}{dx}\hat{\sigma}_y\,.
\end{equation}

Suppose that the SOI term is small and treat it as a perturbation to study the scattering of electrons incident from the left and right reservoirs. Unperturbed wave functions are
\begin{equation}
 \psi_{\lambda,k_x,k_y}^{(0)}=\dfrac{1}{\sqrt{2}}\binom{\lambda \chi}{1}e^{i(k_x x+k_y y)}\,,
\label{psi_0}
\end{equation}
where $\chi=(k_y+i k_x)/k$. The scattered wave functions are easily found in the Born approximation. We are interested mostly in the wave functions outside the SOI strip. For $x>l$ one finds
\begin{equation}
 \psi_{\lambda,k_x,k_y}^{(1)}\simeq -\dfrac{i m}{\hbar^2|k_x|}\left\{
\begin{aligned}
&\tilde{\alpha}_0 \lambda k \psi_{\lambda ,k_x,k_y}^{(0)}, & k_x>0\,,\\
&\tilde{\alpha}_{2|k_x|} \lambda k_y \chi \psi_{\lambda,-|k_x|,k_y}^{(0)}, & k_x<0\,,
\end{aligned}
\right.
\label{perturb1}
\end{equation}
where $\tilde{\alpha}_0\!=\!\int_{-\infty}^{\infty}dx\alpha(x)$ and $\tilde{\alpha}_{2k_x}\!=\!\int_{-\infty}^{\infty}dx\alpha(x)\exp(2ik_xx)$.

Using these wave functions we get simple expressions for the spin density and spin currents averaged over the equilibrium distribution function. The results are as follows. The spin density equals zero, $\Delta \mathbf{S}=0$. It is curious that in each state $\psi_{\lambda,\mathbf{k}}$, the spin density is perturbed, but summing over $\lambda$ nullifies all spin components excluding the $x$ component. The latter is an odd function of $k_y$:
\begin{equation}
\Delta S_{x}(\mathbf{k})\simeq\dfrac{2mk_y}{\hbar |k_x|}\mathrm{Im}\left[\tilde{\alpha}_{2|k_x|}e^{2i|k_x|x}\right]\,. 
\label{spin_perturb}
\end{equation}
This spin density perturbation is produced only by backscattered electrons. Transmitting electrons do not contribute to the spin density, in the first approximation. The integration of $\Delta S_{x}(\mathbf{k})$ with respect to $\mathbf{k}$ with the equilibrium distribution function results in the zero $\Delta \mathbf{S}$.

The tangential spin current in each electron state is simply a product of the spin density and tangential velocity. So the only nonzero component of the spin current, after  summing over $\lambda$, is $\mathcal{J}_y^x(\mathbf{k})=(\hbar k_y/m)\Delta S_x(\mathbf{k})$, which is an even function of $k_y$. Finally the total spin current is 
\begin{equation}
\mathcal{J}_y^x\approx \dfrac{2}{3\pi^2}\int_{k_c}^{k_F}\dfrac{dk}{k}\left(k_F^2-k^2\right)^{3/2}\mathrm{Im}\left[\tilde{\alpha}_{2k}e^{2ikx}\right]\,,
\end{equation}
where $k_c$ is a low-energy cutoff caused by the violation of the Born approximation. Integration leads to resultы qualitatively similar to those shown in Fig.~\ref{Jyxz}(a). The $z$ component of the spin current appears in the second approximation.

The mechanism of the equilibrium spin current in the normal region can be interpreted as follows. Electrons of the normal 2DEG, while they are scattered back by the SOI region, suffer the action of an effective magnetic field which forces their spin to precess. This field is directed normally to the wave vector of the backscattered wave. It is easily seen that due to the precession the spin vectors of both polarizations of incident electrons get equal augmentations of the $x$ projection while the variations of other projections are of opposed signs. Electrons incident with opposed $k_y$ produce the spin projections on the $x$ axis of opposed sign, so that the total spin density is zero. However, the spin currents produced by electrons with opposed $k_y$ are added.

In conclusion, the edge spin currents exist in 2D systems with spatially nonuniform SOI at thermodynamic equilibrium. These currents exist even in the normal 2DEG if it borders with a SOI region. The spin density flows tangentially to the border within a strip of the width, which is determined by the characteristic wave vector of the SOI, the spins being polarized in the plane perpendicular to the border line. These spin currents are proportional to the SOI constant $\alpha$, and are much higher than the equilibrium spin current in an infinite 2D SOI system, since the latter is proportional to $\alpha^3$. The edge spin currents originate from the spin-dependent scattering of electrons from the nonuniform SOI region.

\acknowledgments
We dedicate this paper to Prof. V.B. Sandomirsky on the occasion of his 80th birthday. This work was supported by Russian Foundation for Basic Research (project No~08-02-00777) and Russian Academy of Sciences (programs ``Quantum Nanostructures''and ``Strongly Correlated Electrons in Semiconductors, Metals, Superconductors, and Magnetic Materials'').


\begin{thebibliography}{30}
\bibitem{Wolf}
S.A.~Wolf, D.D.~Awschalom, R.A.~Buhrman, J.M.~Dau\-ghton, S. von Moln\'{a}r, M.L.~Roukes, A.Y.~Chtchelkanova, and D.M.~Treger, Science \textbf{294}, 1488 (2001).

\bibitem{Zutic} 
I.~\v{Z}uti\'{c}, J.~Fabian, and S. Das Sarma, Rev. Mod. Phys. \textbf{76}, 323 (2004).

\bibitem{Murakami} 
S.~Murakami, N.~Nagaosa, and S.C.~Zhang, Science \textbf{301}, 1348 (2003).

\bibitem{Sinova}
J.~Sinova, D.~Culcer, Q.~Niu, N.~A.~Sinitsyn, T.~Jungwirth, and A.~H.~MacDonald, Phys. Rev. Lett. \textbf{92}, 126603 (2004).

\bibitem{Rashba1}
E.I.~Rashba, Phys. Rev. B \textbf{68}, 241315(R) (2003).

\bibitem{Pareek}
T.P.~Pareek, Phys. Rev. Lett. \textbf{92}, 076601 (2004).

\bibitem{Rashba2}
E.I.~Rashba, Phys. Rev. B \textbf{70}, 161201(R) (2004).

\bibitem{Burkov}
A.A.~Burkov, A.S.~N\'u\~nez, and A.H.~MacDonald, Phys. Rev. B \textbf{70}, 155308 (2004).

\bibitem{Shi} 
J.~Shi, P.~Zhang, D.~Xiao, Q.~Niu, Phys. Rev. Lett. \textbf{96}, 076604 (2006).

\bibitem{Sonin1}
E.B.~Sonin, Phys. Rev. B \textbf{76}, 033306 (2007).

\bibitem{Sun}
Q.-F.~Sun, X.C.~Xie, and J.~Wang, Phys. Rev. Lett. \textbf{98}, 196801 (2007); Q.-F.~Sun, X.C.~Xie, and J.~Wang, Phys. Rev. B \textbf{77}, 035327 (2008).

\bibitem{Sonin2}
E.B.~Sonin, Phys. Rev. Lett. \textbf{99}, 266602 (2007). 

\bibitem{n1}
Spin current in a hybrid (normal-SOI) ring with SOI sector was studied in Ref.~\onlinecite{Sun}. We consider a qualitatively different situation of 2D system, in which two spatial components of the spin current are important. 

\bibitem{Rokhinson}
L.P.~Rokhinson, L.N.~Pfeiffer, and K.W.~West, Phys. Rev. Lett. \textbf{96}, 156602 (2006).

\bibitem{Sablikov}
V.A.~Sablikov and Yu.Ya.~Tkach, Phys. Rev. B \textbf{76}, 245321 (2007).

\bibitem{n2}
Note that no edge quantum states exist in the system under consideration.

\end{thebibliography}
\end{document}